\documentclass{bmcart}

\usepackage[utf8]{inputenc} 

\def\includegraphics{}

\startlocaldefs
\usepackage{color}
\usepackage{soul}
\usepackage{endnotes}

\usepackage{graphicx}
\usepackage{changes}
\endlocaldefs

\begin{document}

\begin{frontmatter}
\begin{fmbox}
\dochead{Research}

\title{Predicting the temporal activity patterns of new venues}

\author[
   addressref={aff1},                   		 
   corref={aff1},                       			 
   email={krittika.dsilva@cl.cam.ac.uk}        
]{\inits{KD}\fnm{Krittika} \snm{D'Silva}}
\author[
   addressref={aff2},
   email={noulas@nyu.edu}
]{\inits{AN}\fnm{Anastasios} \snm{Noulas}} 
\author[
   addressref={aff3,aff4},	
   email={m.musolesi@ucl.ac.uk}
]{\inits{MM}\fnm{Mirco} \snm{Musolesi}} 
\author[
   addressref={aff1,aff4},
   email={cecilia.mascolo@cl.cam.ac.uk}
]{\inits{CM}\fnm{Cecilia} \snm{Mascolo}} 
\author[
   addressref={aff5},
   email={max@foursquare.com}
]{\inits{MS}\fnm{Max} \snm{Sklar}} 
 
\address[id=aff1]{                           						 
  \orgname{Department of Computer Science, Cambridge}, 	 
  \street{15 J J Thomson Ave},                      
  \postcode{CB3 0FD}                                					 
  \city{Cambridge},                              					 
  \cny{UK}                                    						 
}
\address[id=aff2]{
  \orgname{Center for Data Science, New York University},
  \street{5th Avenue},
  \postcode{10011}
  \city{New York},
  \cny{USA}
}
\address[id=aff3]{
  \orgname{Department of Geography, University College London},
  \street{Gower Street},
  \postcode{WC1E 6BT}
  \city{London},
  \cny{UK}
}
 \address[id=aff4]{
  \orgname{The Alan Turing Institute}, 
  \city{London},
  \cny{UK}
}
\address[id=aff5]{
  \orgname{Foursquare Labs}, 
  \city{New York},
  \cny{USA}
}
\end{fmbox} 

\begin{abstractbox}
\begin{abstract}
Estimating revenue and business demand of a newly opened venue is paramount as these early stages often involve critical decisions such as first rounds of staffing and resource allocation. Traditionally, this estimation has been performed through coarse-grained measures such as observing numbers in local venues or venues at similar places (e.g., coffee shops around another station in the same city).
The advent of crowdsourced data from devices and services carried by individuals on a daily basis has opened up the possibility of performing better predictions of temporal visitation patterns for locations and venues.  
In this paper, using mobility data from Foursquare, a location-centric platform, we treat venue categories as proxies for urban activities and analyze how they become popular over time. 
The main contribution of this work is a prediction framework able to use characteristic temporal signatures of places together with k-nearest neighbor metrics capturing similarities among urban regions, to forecast weekly popularity dynamics of a new venue establishment in a city neighborhood. We further show how we are able to forecast the popularity of the new venue after one month following its opening by using locality and temporal similarity as features. For the evaluation of our approach we focus on London. We show that temporally similar areas of the city can be successfully used as inputs of predictions of the visit patterns of new venues, with an improvement of 41\% compared to a random selection of wards as a training set for the prediction task. 
We apply these concepts of temporally similar areas and locality to the real-time predictions related to new venues and show that these features can effectively be used to predict the future trends of a venue. 
Our findings have the potential to impact the design of location-based technologies and decisions made by new business owners.
\end{abstract}


\end{abstractbox} 
\end{frontmatter}

\section*{Content}
\section{Introduction}
Cities are complex systems that constantly change over time. From city dwellers that commute to work on a weekday morning to visitors who arrive in town for business or leisure, the urban landscape is transforming at a fast pace. 
The way in which city neighborhoods become popular over time has been a fundamental area of study in traditional urban studies literature as it is critical to city governance~\cite{kaiser1995urban, batty2007cities}. The rise of mobile technologies and collective sensing in the last decade has contributed to the generation of large datasets that describe activity dynamics in cities and has created new opportunities for research in the area; for example, a number of works have proposed the use of cellular data to understand collective mobility dynamics and inform planning decisions~\cite{reades2007cellular, calabrese2011real, ratti2006mobile, becker2013human, jiang2016timegeo}. Beyond cellular data, the increasing popularity of services like Twitter and Foursquare has yielded new inputs for capturing the \textit{heartbeat} of a city~\cite{francca2016visualizing, silva2013comparison, zheng2014urban}. Further, trends in credit card transactions have also been used to study human patterns across space and time~\cite{creditCardDynamics}. 

On the venue level, temporal dynamics and the spatial configuration of urban activities has helped decide where to geographically place new retail facilities ~\cite{jensen2006network, karamshuk2013geo} as well as to power mobile applications such as local search~\cite{shaw2013learning} by exploiting place temporal dynamics. Nevertheless, little work has looked at predicting what happens after a new venue opens in a city neighborhood; \textit{will it become popular?} and moreover, \textit{at which times of the week should the owner of a retail facility expect high volumes of customer traffic?} This information is important during the early stages of a new business when staffing levels must be decided, supplies bought, and opening times established.

In this paper, starting from the premise that mobility in a city is driven by local urban activities, we provide an analytical framework that captures the popularity dynamics of urban neighborhoods. We then exploit these temporal patterns across areas to predict the popularity dynamics of newly established venues.
The primary data input we use for our study is a longitudinal dataset from location-based service Foursquare~\endnote{www.foursquare.com} describing mobility in terms of user \textit{check-ins} at public venues in the city of London. 
Our approach can be summarized as follows:

\begin{itemize} 
\item \textbf{Temporal characterization of urban activities across regions:} 
First, we show how the temporal profile of an area in terms of the number of mobile users that visit over the course of a week varies significantly from neighborhood to neighborhood. These temporal profiles are shaped by the daily and weekly circadian rhythms of moving populations as well as their choice of specific urban activities at key times. Further, we demonstrate how the popularity dynamics of venue categories give rise to the temporal patterns of the urban areas that contain them, highlighting how urban activities and population levels at a neighborhood are inherently interconnected temporal processes. 
\item \textbf{Predicting the popularity dynamics of newly established venues:}
Next, we harness the temporal popularity patterns of venue categories to predict the weekly temporal profile of a newly established venue. 
Our approach is based on two key principles. First, we exploit the fact that despite their differences, neighborhoods in a city can exhibit a high degree of temporal \textit{synchronicity}, even if they are geographically located far from one another. For example, the city of London includes neighborhoods such as SoHo and Camden Town, both of which attract a young late-night crowd. These shared traits could mean these neighborhoods are likely to become popular at similar times. Second, we make use of the principle of \textit{locality} that suggests that the temporal profile of venues in the same neighborhood are highly correlated as individual movements in a city tend to be constrained by distance. We combine 
the properties of \textit{synchronicity} and \textit{locality} in a novel k-nearest neighbor model 
to determine similarities between neighborhoods in cities and use this understanding to predict the characteristic demand curve of a new venue. We use classic Gaussian Processes along with a k-nearest neighbor approach to build a model which accurately predicts a new venue's popularity temporal profile. Our results perform significantly better than our random baseline, decreasing the normalized root mean square error  by 41\%.

\item \textbf{Real time demand prediction of new venues:} Finally, given a new venue, our goal becomes to predict how the demand changes as that venue matures over time. With each progressive month of growth for a new venue, it is to be expected that some venues flourish and grow in popularity over time while others may be less successful and see their demand decline over subsequent months. Although sparsity can be an issue when working at a fine spatial granularity, making the formulation of a regression problem challenging, we show that it is possible to improve the prediction of expected levels of visits for the next time step by using an approach based on neighborhood \textit{synchronicity} and \textit{locality}. We train a Gaussian Process model on the month to month trends of check-ins to other venues and wards and use these as inputs to forecast the popularity of a new venue. In doing so, we incorporate recent changes in urban mobility, which could arise for instance due to the presence of new events nearby or other anomalies such as transport disruptions. 
\end{itemize}

Our work enables a fine-grained dynamic estimation of activity for new venues.  
Obtaining analytics in this context can help can business owners predict demand in dynamics for their business
and therefore plan better the provision of services to their customers. 
 
The remainder of the paper is organized as follows. In Section 2, we motivate our work and introduce the related work in the area. Section 3 gives an overview of our approach and Section 4 introduces a formalization of our framework. Section 5 reports on our temporal analysis of venues and neighborhoods in London. In Section 6, we describe a method for predicting the characteristic temporal profiles of a set of new venues using a batch-learning approach, whereas in Section 7 we present an analysis of the real-time extension of this approach. We conclude the paper with Section 8 discussing our results and Section 9 highlighting possible future work.

\section{Related Work}
The ubiquity of GPS sensors on mobile devices as well as the introduction of the mobile web have been game 
changers with respect to the scales and types of data available about human mobility. 
The rise of services such as Foursquare and more generally applications that rely on geo-tagging technologies  
 (e.g. Twitter, Instagram, Flickr) combined with the accessibility to the corresponding APIs have offered novel views to collective mobility activity in cities~\cite{noulas2011empirical, pan2013crowd}. 
This has led to more granular representations of urban activities 
across space and time~\cite{francca2016visualizing, zheng2014urban, silva2013comparison}, 
and has been used to characterize cities in terms of their urban growth patterns~\cite{daggitt2016tracking} and cultural boundaries in terms of their culinary patterns~\cite{silva2014you}. Additionally, such location-based technologies have significantly improved the quality of 
experience for mobile users as they navigate the city. Foursquare's data
science team exploited the weekly temporal visitation patterns of venues to power its local search engine~\cite{shaw2013learning}.  Google Maps recently incorporated 
the feature \textit{popular times} that appears when search results about places are shown to users~\cite{googletimes}, while Facebook launched its in-house place search service~\cite{facebookplaces}. 
The commercialization potential of such services has naturally expanded beyond the realm of location-based
technologies and has contributed to coining the term \textit{location intelligence} when referring to business
intelligence relying on geospatial data. In this direction, a number of works have appeared on retail optimization in cities, including the identification of the best location to open a new shop~\cite{jensen2006network, karamshuk2013geo} 
or the ranking of areas according to their real estate value~\cite{fu2014sparse}. 
 
Before datasets from location-based services became available, those collected from cellular networks paved the way
for understanding the collective dynamics of urban activities. 
Ratti et al. in~\cite{ratti2006mobile} present one of the first works that
demonstrates how urban landscapes transform in real time as populations move around the city. Beyond 
dynamic visualizations, the authors in~\cite{reades2007cellular, calabrese2011real} characterize in statistical terms
urban activity variations and provide interpretation on the observed patterns in terms of the underlying urban
activities, such as transport and residential land uses, which drive population volumes regionally. Becker et al. 
in~\cite{becker2013human} use cellular data to characterize mobility trends across different metropolitan 
areas, while the authors in~\cite{jiang2016timegeo} propose using cellular data as an alternative to 
travel surveys so that more accurate spatio-temporal representations of mobility flows are obtained. 
Urban transport data has often been another source for capturing city dynamics~\cite{roth2011structure, smith2013finger}.

While in many of the works mentioned above, information on the temporal visitation patterns of users to locations 
has been used as input, none has looked at predicting the temporal signatures of visits to venues per se. 
This is where the primary novelty of the present paper lies. Considering the temporal patterns of user visits at newly established venues as our main prediction task we are hoping to offer
new insights on location-based analytics 
for business owners that would empower them to make more informed choices on staffing, provision of goods, resources in the early days of their new business. We also envision the use of similar approaches to inform a number of tasks that are applicable in the urban domain. These may include the spatial deployment of taxi fleets and pooling services~\cite{balan2011real, santi2014quantifying} or the allocation of police or ambulance resources~\cite{rajagopalan2008multiperiod,sheu2007emergency}.

\section{Our Approach At A Glance} 
In this section, we describe how our approach to the prediction of visitation patterns to new venues harnesses temporal similarities of urban neighborhoods. Our analysis revolves around the concept of a characteristic weekly temporal profile, a time series representing typical changes in demand of a given entity over the course of a week. We explore characteristic profiles of venues, which fluctuate based on user mobility patterns, and neighborhoods that are composed of venue profiles.

We begin with the premise that venue categories have different characteristic weekly temporal profiles. These profiles represent variations in demand based on a user's propensity to visit that category at a given hour of the week.  For example, the category of Travel \& Transport is likely to correlate with changes in rush hour traffic while Food could instead be dependent on typical meal times. Different neighborhoods in a city have characteristic weekly temporal profiles which are made up of contributions from different categories. We posit that neighborhoods which have similar temporal profiles or similar contributions of venues to their temporal profile could be predictors for each other. We apply this idea towards the analysis of new venues: \textit{given a new venue in a given neighborhood in a city, can we use the the demand profile of venues in temporally similar wards as predictors for our new venue of interest?} For new venues opening up in a city, often no prior information is known about the expected popularity or demand dynamics. The ability to approximate and better understand these metrics can be crucial for the success of a new business owner.
 
This characteristic temporal curve provides a static representation of the typical changes in demand of a venue over the course of a week. We build upon our analysis of new venues by dynamically predicting how the demand of a venue will change. Starting from one week after a venue has opened, we show that we can use data from venues with similar characteristic curves to more accurately predict the changes in demand at the next time step.

\section{Notation and Definitions}
\subsection{Dataset}
Online Location-based Social Networks (LBSNs) have recently experienced a surge in popularity, attracting millions of users around the world. The widespread adoption of these services in addition to location-sensing mobile devices has created a wealth of data about the mobility of humans in cities. Foursquare, a popular location-centric media platform, enables users to check into different locations and share that information with their friend group. As of August 2015, Foursquare had more than 50 million active users and more than 10 billion check-ins~\cite{foursquarenumbers}. 

For our work we use a longitudinal dataset describing urban mobility and activity patterns in Greater London that spans three years and millions of check-ins. For each venue, we have the following information: geographic coordinates, specific and general category, creation date, total number of check-ins, and number of unique visitors. The specific and general categories fall within Foursquare's API of hierarchical categories. A full list of the categories can be found by querying the Foursquare API\endnote{https://developer.foursquare.com/categorytree}. General categories are overarching groups to one of which each specific category is assigned. Examples of general categories could include \textit{Food} or \textit{Travel \& Transport} while examples of specific categories could be \textit{Chinese Restaurants} or \textit{Italian Restaurants}, which both aptly fall under the category of \textit{Food}. In addition to data about venues, the dataset also contains \textit{transitions} within London. A transition is defined as a pair of check-ins by an anonymous user to two different venues within the span of three hours. A transition is identified by a start time, end time, source venue, and destination venue. Our dataset includes 18,018 venues and 4,000,040 transitions for Greater London.  The dataset comprises of check-ins from December 2010 to December 2013. This dataset was obtained through a collaboration with Foursquare.

\subsection{Formalization}
In this section, we introduce a formalization of our model. Electoral wards are the main building blocks of administrative geography in the United Kingdom; Greater London consists of 649 electoral wards and these spatial units uniquely identify London boroughs ~\cite{nationalwardstats}. We use wards $w \in \mathbf{W}$ as a means of subdividing Greater London. We also consider venues $v \in \mathbf{V} $. A venue has a precise geographic location in a ward. 
A venue $v$ is represented with a tuple $v = <loc, g, s>$ where $loc$ is the geographic location of the venue,  $g$ its general category and $s$ is its specific category.

We define a \textit{time interval} $t$ as the interval $[t\Delta, (t+1)\Delta]$ of duration $\Delta$. For example the time interval $t=0$ indicates the interval $[0, \Delta]$, the time interval $t=1$ indicates the time interval $[\Delta, 2\Delta]$ and so on. In our work, each time interval represents distinct hours and do not overlap.

\textbf{Definition 1: Temporal Profile of a Ward.} Similarly we define the temporal profile of a ward $w$ in an interval $[0,T]$ as the following sequence (i.e, time series):

\begin{equation}
C^{w}[0,T] = \{c_{t}^{w}\} \quad \mathrm{with} \quad t = {0, 1, \dots T-1}
\end{equation}

\noindent where $c_{t}^{w}$ is the \textit{total} number of check-ins in the ward $w$ during the time interval $t$.

\textbf{Definition 2: Temporal Profile of a Venue.}  We define the \textit{temporal profile of a venue} $v$ in an interval $[0,T]$ as the following sequence (i.e, time series):

\begin{equation}
C^{v}[0,T] = \{c_{t}^{v}\} \quad \mathrm{with} \quad t = {0, 1, \dots T-1}
\end{equation}

\noindent where $c_{t}^{v}$ is the \textit{total} number of check-ins to venue $v$ during the time interval $t$.

\textbf{Definition 3: Aggregate Temporal Profile of Venues of a Generic (Specific) Category in a Ward.} We then define $\mathbf{V}_{g,w}$ as the set of the venues of \textit{generic} category $g$ in a ward $w$. 
Similarly, we define $\mathbf{V}_{s,w}$ as the set of the venues of \textit{specific} category $s$ in a ward $w$.

Therefore, the \textit{aggregate temporal profile} of venues of \textit{generic} category $g$ in a ward $w$ in a time interval $[0,T]$ is defined as the following sequence (i.e, time series):

\begin{equation}
C^{{V}_{g,w}}[0,T] = \{c_{t}^{g,w}\} \quad \mathrm{with} \quad t = {0, 1, \dots T-1}  
\end{equation}

\noindent where $c_{t}^{g,w}$ is the \textit{total} number of check-ins to venues of general category $g$ in the ward $w$ during the time interval $t$. 
The temporal profile of venues of a specific category in a ward can be defined in a similar way.

 \section{Temporal Patterns of Mobile User Activity}
Having formally defined the concept of temporal profiles, in this section, we discuss temporal trends of wards within Greater London and demonstrate how the composition of those wards plays a crucial role in creating a characteristic profile for that ward. We begin with an examination of all wards at one particular point in the day, highlighting that different categories dominate different wards at any given point. We then analyze more closely the characteristic temporal profile of two wards in London and discuss how their different category types contribute to their different overall profiles. We then quantify the similarity between the overall temporal profile of the 15 most popular wards in London and discuss how similarity in temporal visitation patterns could inform predictions for the temporal profile of a new venue.

\subsection{Regional temporal activity patterns}
\begin{figure}[h!]
\includegraphics[scale=0.35] {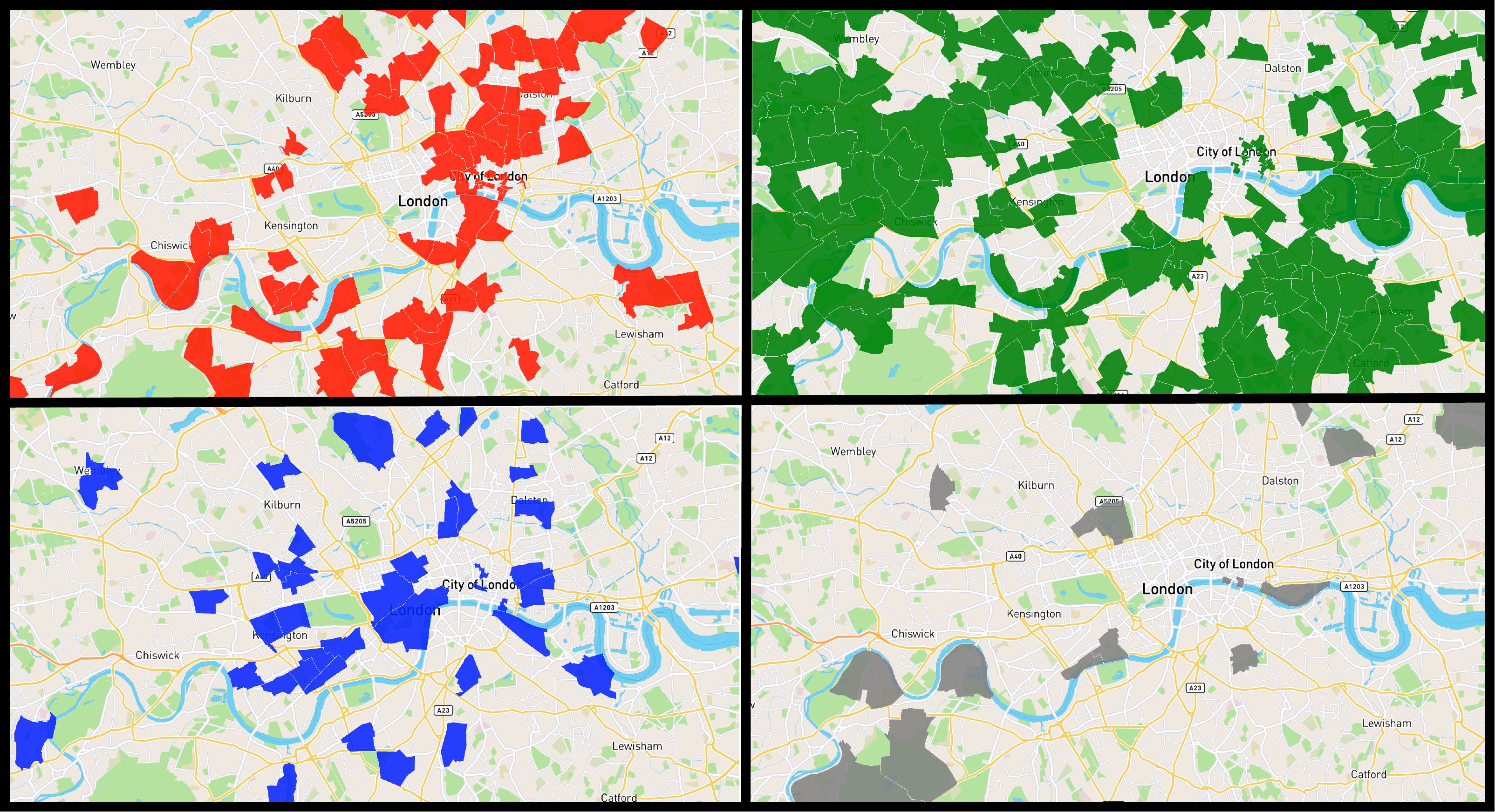}
\caption{\csentence{Venue category popularity.} 
The most popular category in each ward at 17:00. \textit{Nightlife Spots} are represented in red, \textit{Travel \& Transport} in green, \textit{Food} in blue, and \textit{Outdoors \& Recreation} in grey.}
\label{ward_categories}
 \end{figure}

Looking broadly at all wards within the city of London, we choose one hour of the day to highlight the idea that the popularity patterns of different neighborhoods can be dominated by different categories. Figure~\ref{ward_categories} shows the most popular category in each ward in London in the time interval $t = 17$ where $\Delta = 1$ (i.e., between 5 to 6pm). For certain wards, this time of the day could be dominated by transport traffic as individuals commute towards home. For others, the most significant contributor could be nightlife, as individuals head to the pub for an evening drink. Similarities in the contribution of different categories to the overall temporal trend of a ward could be an indication that those wards attract individuals with similar demographics or have similar characteristics. An analysis of these similarities can be harnessed to better model, characterize, and profile different wards in a city.

\begin{figure}[h!]
\includegraphics[scale=0.55] {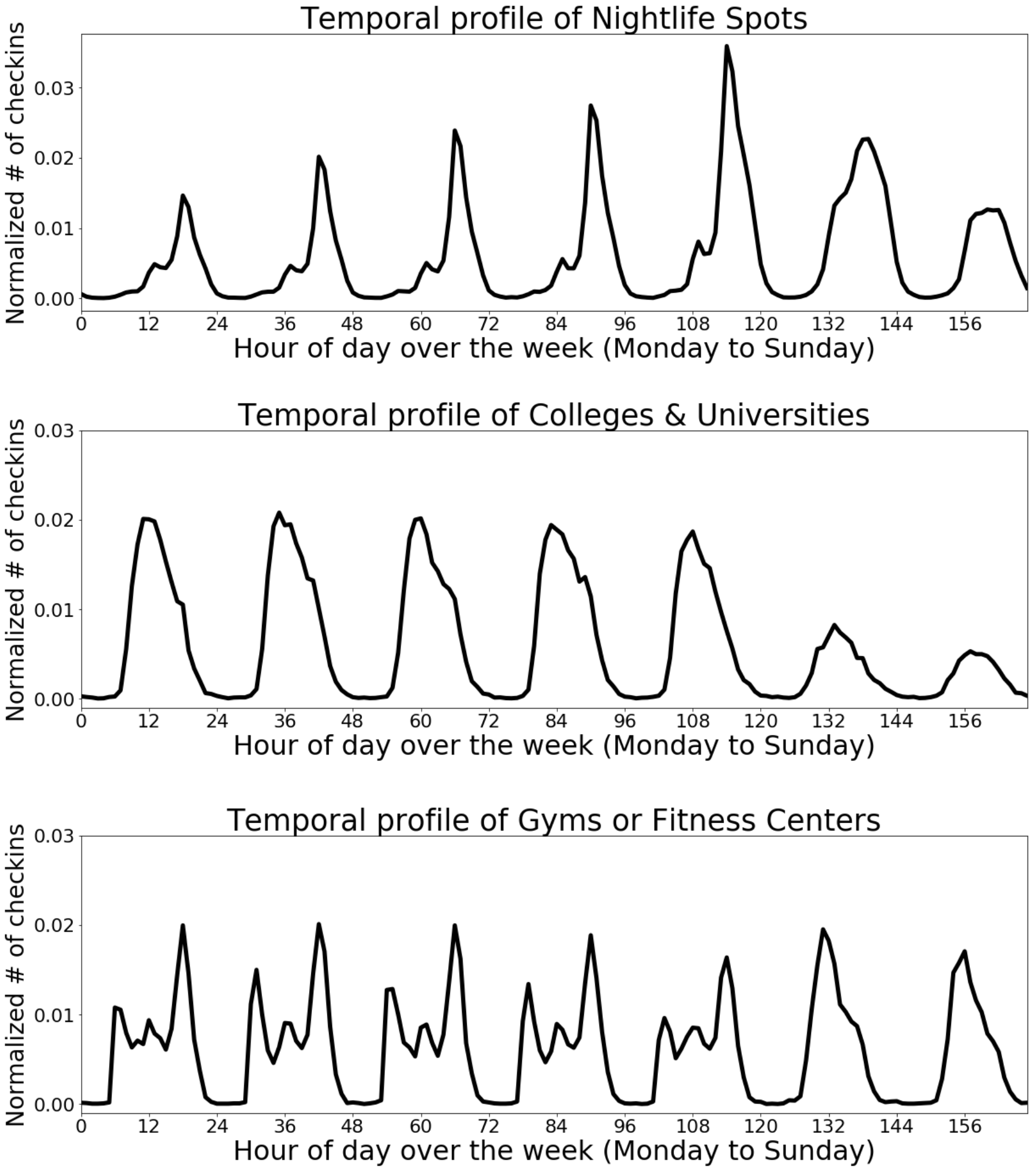} \
\caption{\csentence{Temporal profiles.} 
Normalized temporal profile of different categories of venues.}
\label{profiles_category}
\end{figure}

Looking more closely at category types, Figure~\ref{profiles_category} presents the characteristic temporal profile of three categories: \textit{Nightlife Spots}, \textit{Colleges \& Universities}, and \textit{Gyms} or \textit{Fitness Centers}. Each profile is a direct function of a users's propensity to visit at a given hour of the day and day of the week. The profiles of different venue categories in a ward establish the overall profile of that venue. A close examination of different wards within the city of London and of the categories which make up those wards present a number of interesting insights on how those vary in terms of their temporal visitation patterns. 

\begin{figure}[h!]
\includegraphics[scale=0.55] {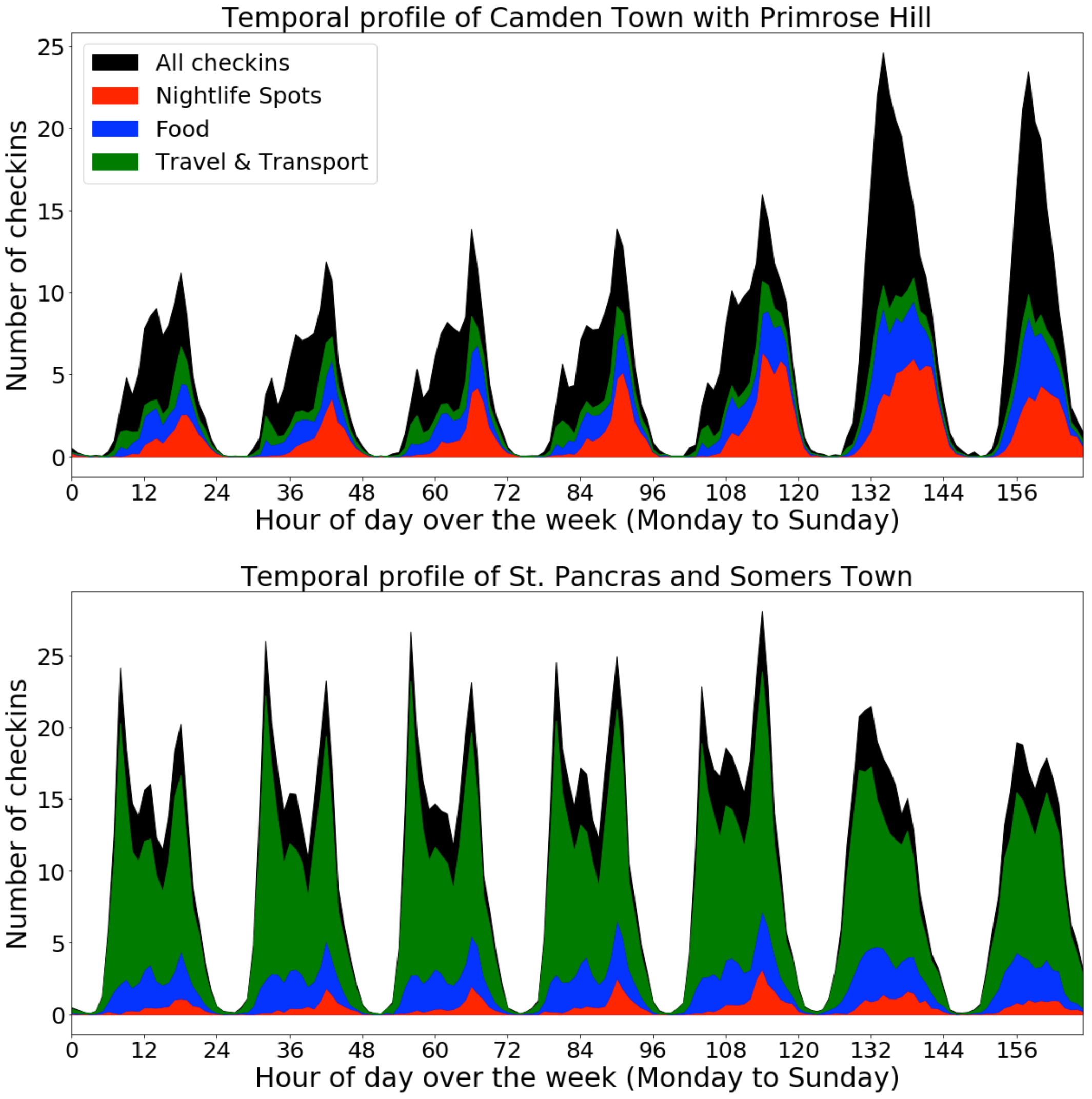}
\caption{\csentence{Ward temporal profile.} 
Daily temporal profiles and category breakdown of St. Pancras \& Somers Town and Camden Town with Primrose Hill, two contrasting wards in London.}
\label{ward_profiles}
\end{figure}
 
 \begin{figure}[h!]
\includegraphics[scale=0.55] {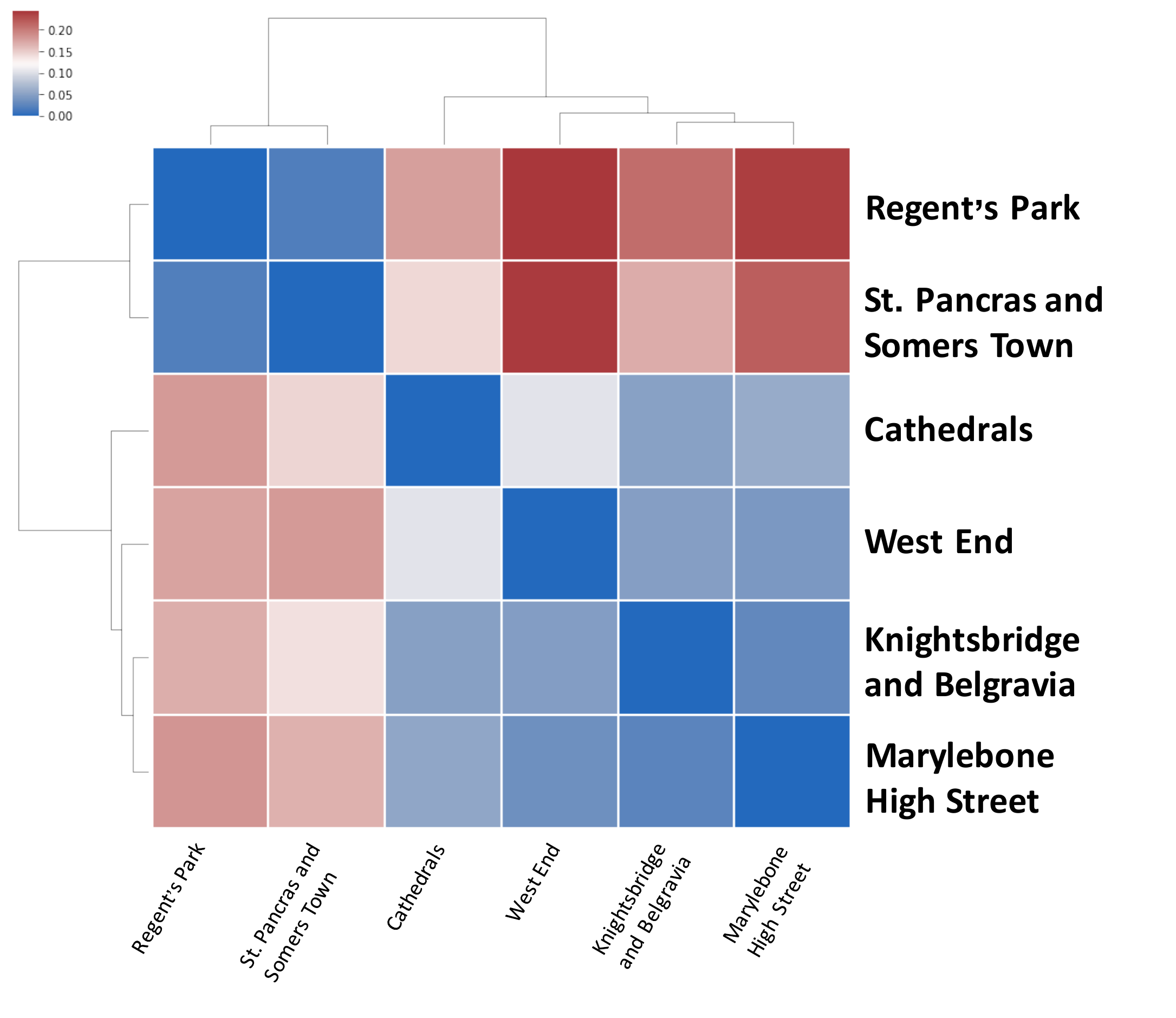}
\caption{\csentence{Temporal similarity of wards.} 
J-S divergence of the characteristic weekly temporal profile of the 15 most popular wards. Smaller values signify a smaller divergence and thus more similarity.}
\label{heatmap}
\end{figure}
 
In order to illustrate this let us consider two wards of interest: St. Pancras \& Somers Town, which contains a major transportation hub, offices, and academic institutions, and Camden Town with Primrose Hill, which contains a variety of venues and tourist attractions. Figure~\ref{ward_profiles} shows the average number of check-ins in each ward for each hour of the day over the course of one week, aggregating across
a number of weeks. This signal creates a characteristic temporal profile which acts as a temporal signature for the ward. The overall signal, shown in black, is different for these two wards. The number of check-ins at Camden Town steadily increases over the course of the day while the number check-ins at St. Pancras has two large peaks, one in the morning and another in the evening. Examining the three main categories (\textit{Food}, \textit{Travel \& Transport}, and \textit{Nightlife Spots}) that characterize these two wards can help to better understand this observation. Camden Town has significant contributions from \textit{Nightlife} venues which gradually increases over the course of a day. Conversely, St. Pancras is dominated by \textit{Travel \& Transport}, causing the overall temporal profile of the ward to peak at rush hour. These trends suggest that Camden Town is likely a more youth dominated area while St. Pancras is a hub for commuters or travelers, as they actually are ~\cite{londonatlas}. 

\subsection{Utilizing similarities in visitation patterns} 
\label{sec:similarities_profiles}
Similar observations can be generalized to the rest of the wards in London. Different regions feature different degrees of similarity, an insight which we exploit in Section~\ref{sec:evaluation} to predict the characteristic temporal curves of new venues. We quantify the similarity between two temporal profiles using the Jensen-Shannon divergence (JSD) ~\cite{lin1991divergence}. We use the JSD instead of the Kullback-Leibler divergence (KLD) since the former is a symmetric similarity measure between two functions whereas the latter is not. Our analysis showed that the JSD improved our percent accuracy by 7 to 10\% over the KLD. The JSD between two wards $w_i$ and $w_j$ is calculated as follows: 
\begin{equation}
JSD(C^{w_i}, C^{w_j}) = H\left(\frac{C^{w_i} + C^{w_j} }{2}\right) - \frac{H(C^{w_i}) + H(C^{w_j})}{2}
\end{equation} 

\noindent where $H$ is the Shannon entropy. The JSD provides an \textit{information-theoretic} metric that quantifies how two profiles, which that can be seen as distributions over time, are similar. A low value of the JSD between the temporal profile of two wards represents a high similarity. 

In Figure~\ref{heatmap} we present the Jensen-Shannon divergence between the temporal profiles of 
the 15 most popular wards in London with regards to their total number of check-ins. There is an evident range in similarity between the various wards. 
For instance, Hyde Park is very similar to St. Pancras and Somers Town as both are central travel hubs 
that handle the large commuter flows to local corporate offices and government buildings. 
Other characteristic examples are wards such as St. James's and West End that attract a large tourist population visiting the attractions in the respective areas. These results suggest that similarities in temporal profiles can be useful indicators of similarities in the characteristics of two wards. We aim to use this insight for our overarching goal of predicting the characteristic temporal profile of a new venue. This similarity can be used for prediction. 

First, we explore how the temporal profile of a new venue becomes more and more stationary with each consecutive week and then demonstrate how similar wards can be used for the prediction task.

\begin{figure}[h!]
\includegraphics[scale=0.35] {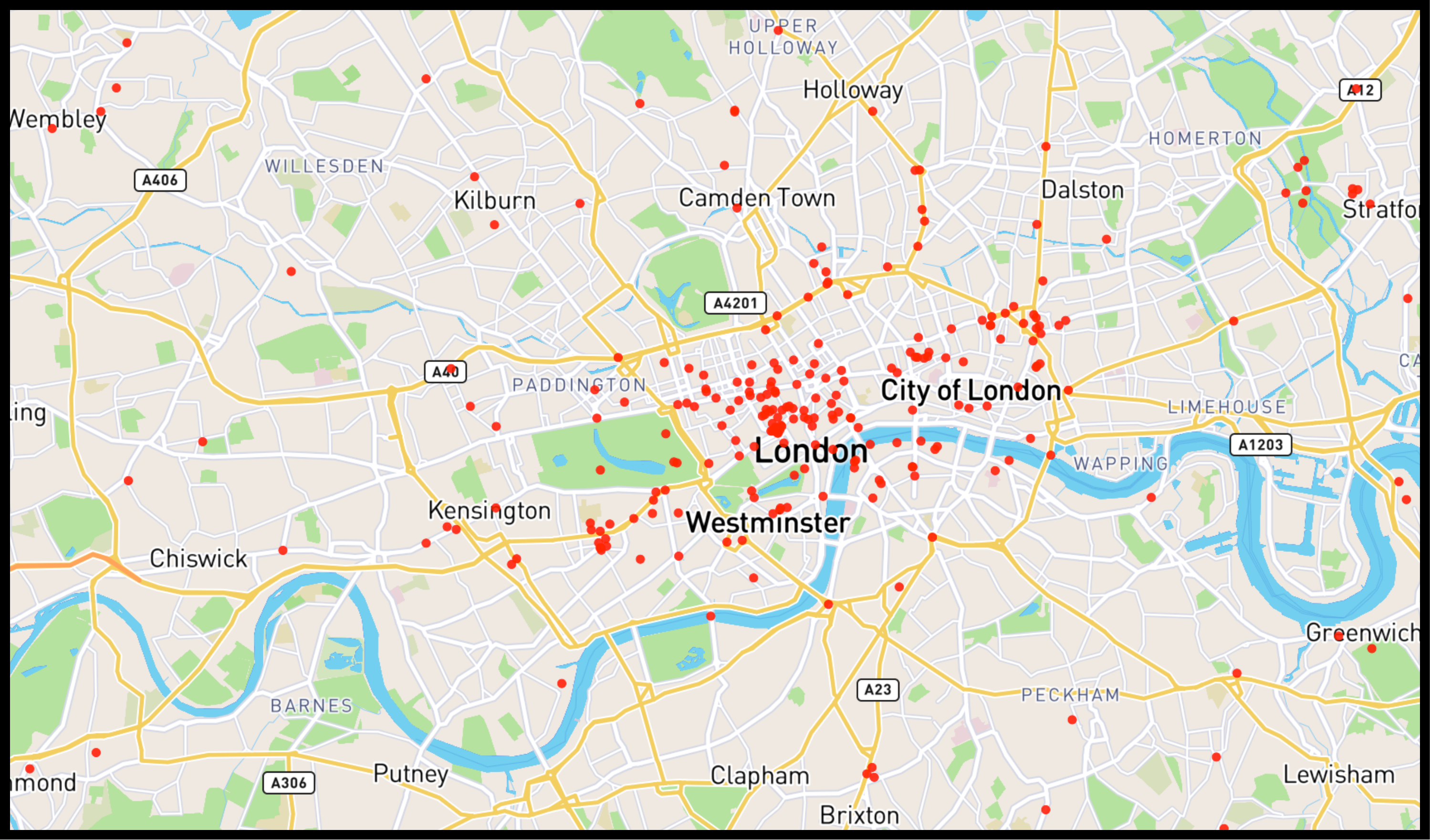}
\caption{\csentence{Set of new venues.} 
Coordinates of the set of 305 new venues in London considered in the study.}
\label{new_venues}
\end{figure}
  
\subsection{Temporal visitation patterns of new venues}
We now focus on the temporal characteristics of new venues. These venues represent an interesting case study 
as upon their launch, unlike existing venues or geographic areas, there is no historic information on their expected popularity patterns over time. We introduce next basic properties of the new venues data that
we use during evaluation. We demonstrate how their temporal profiles converges to a stationary state over time, a process that will let us define the prediction task presented in Section~\ref{sec:evaluation}.  

\textbf{Identification of new venues.} The Foursquare dataset includes a list of all public venues in the city of London. Critically for the present work the \textit{creation time} of every venue is available in the database. The creation 
time refers to the date the venue was crowdsourced by Foursquare users.  
Prior research on Foursquare data has shown that venues added after June 2011 were highly likely (probability above 0.8) to actually be new venues opening in an area rather than existing venues being added to the system for the first time~\cite{daggitt2016tracking}.
We look at all new venues that were added to Foursquare after June 2011 that had at least 100 check-ins. This results in a list of 305 venues which is used for the following analysis. Within this list of newly opened venues, 32\% of the venues fall within the general category of \textit{Food}, 32\% under \textit{Travel \& Transport},  and 8\% under \textit{Nightlife Spots}. These venues were scattered around the city, with their focal point being in center of the city as depicted in Figure~\ref{new_venues}. 

\textbf{Defining a venue's stationary temporal profile.} For each of the new venues in our set, we first examine the total number of checkins at each time step for each week after the venue opened (ie. the weekly temporal profile). To avoid sparsity issues working at this level of granularity, we create a cumulative temporal profile per week, summing the total number of checkins at each time step with each consecutive week. Although the raw values of cumulative sums have little significance, the trend over the course of the week represents the characteristic curve of the venue and indicates the weekly demand trend. We normalize the curve for each week by dividing by the sum of all checkins for the venue, up to the time of observation. With each consecutive week, we expect this curve to show a higher degree of stationarity. We measure the stationarity of this temporal profile over time by calculating the variance of the temporal curve at time $t$ relative to time $t - 1$.

Our data suggests that the temporal profile of a new venue becomes stationary when the value of the variance relative to the prior week is  $\sigma^2 < 2.6 \times 10^{-5}$. On average, this occurs 5 weeks after a venue has opened. Note that we build the profile of a venue considering a week's temporal span. This captures the most essential temporal patterns of activity at a venue, which includes diurnal variations, but also differences between weekends and weekdays.

\section{Predicting the Temporal Signature of New Venues}
\label{sec:evaluation}
Having built an understanding of similarities in the temporal profile of categories and wards, we aim to apply these findings to predict the stationary temporal profile of a new venue. We adopt a k-nearest neighbors approach in which we find the $k$ most temporally similar wards to the ward in which the new venue is located. We look not only at the overall profile of wards as a means of comparison but also at the profile of categories within those wards. The temporal profile of those wards serve as predictors for the new venue and are used to train a Gaussian Process model~\cite{rasmussen2006}.  

\subsection{Discovering area-wide similarities in popularity dynamics} 
\label{sec:model_desc}
\begin{figure}[h!]
\includegraphics[scale=0.35] {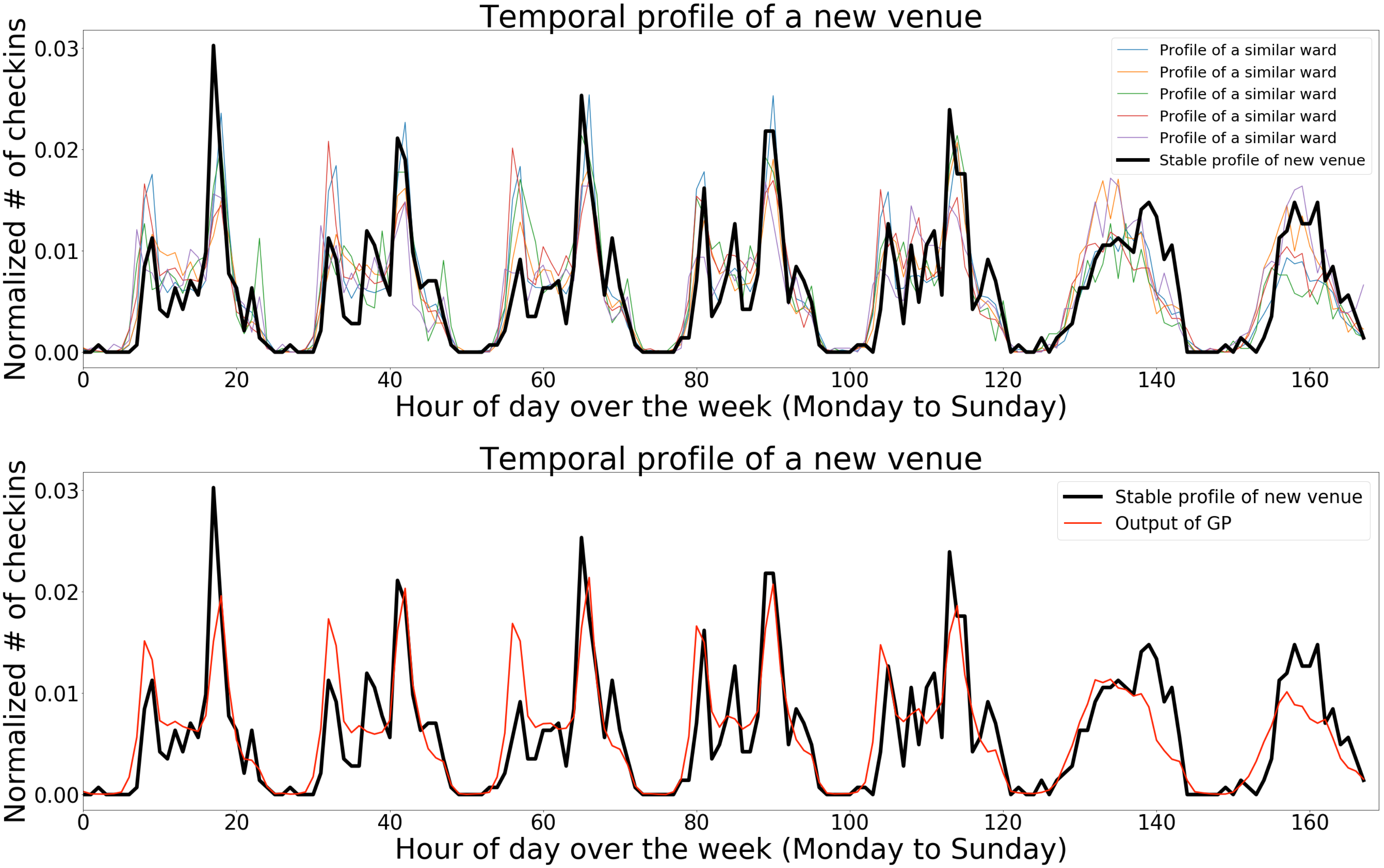}
\caption{\csentence{Stable temporal profiles.} 
Top panel: the normalized stable temporal profile of the new venue with the profile of similar wards. Bottom panel: the output of the GP trained on the similar ward profiles; this serves as a prediction of the profile of the new venue.}
\label{similar_profiles}
\end{figure}

\begin{figure}[h!]
\includegraphics[scale=0.60] {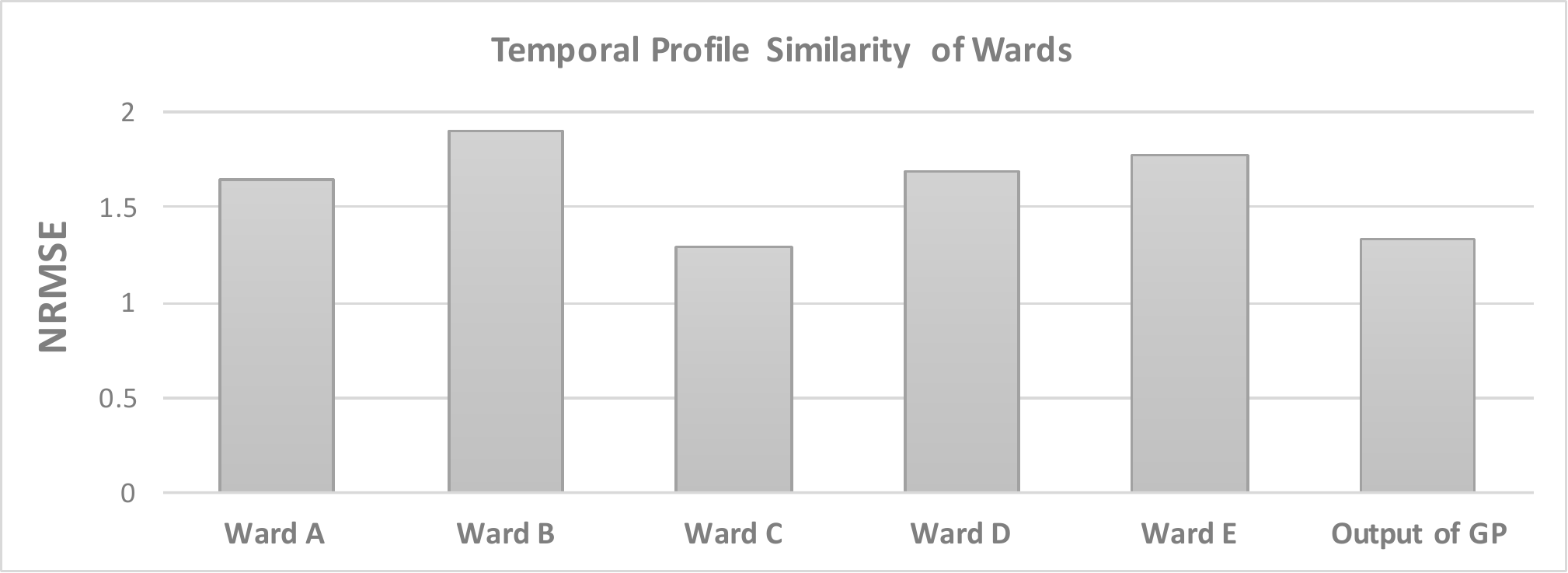}
\caption{\csentence{Profile similarity of different wards.} 
The NRMSE between the stable temporal profile of the new venue and the temporal profile of five similar wards. ``Output of GP" is the NRMSE between the output of the trained GP model and the temporal profile of the new venue.}
\label{wards_nrmse}
\end{figure}

We have seen that similarities in the temporal profiles of wards can be useful indicators of similarities in the characteristics of two wards (Section \ref{sec:similarities_profiles}). We use this idea for our model in which we begin with the basis that two venues of the same category in two different wards are likely to have similar temporal patterns if the overall temporal patterns of their wards are similar. This idea is illustrated in Figure~\ref{similar_profiles} which shows the stable temporal profile of a new venue $v_i$ and the temporal profile of the five most similar wards. Further, Figure~\ref{wards_nrmse} shows the normalized root mean squared error (NRMSE) between the stable profile of a new venue and each of the five similar wards as well as between the profile output from the GP. The figure demonstrates that GP predictors provides a better prediction with respect to simply using the profiles of similar wards.

For a given new venue $v_i$, our methodology to predict its temporal profile is as follows. For clarity, we will describe an example in which we assume $v_i$ is an Italian restaurant called \textit{The Meaning of Life} in ward 42.  
\begin{enumerate}
  \item Determine the general category, specific category, and ward of that venue. For our example, the general category is \textit{Food}, the specific category is \textit{Italian restaurant}, and the ward is 42.
  \item Determine the temporal profile of the ward for the general category of interest. In this example, we would determine the overall temporal profile of \textit{Food} venues in ward 42. Formally, we determine $C^{{V}_{g,w}}[0,T]$ where $T=168$.
  \item Determine the $N$ most similar wards. For all other wards in the city, compare their general category's temporal profile to that of our ward of interest and determine the $N$ most similar wards where similarity is defined as $JSD(C^{v}, C^{w})  \quad v \neq w$. 
This is referred to as the set of \textit{temporally similar wards}. For our example, this would entail finding the $N$ wards whose Food temporal profile is most similar to that of ward 42. 
  \item Calculate the specific temporal profile for each ward in the set of \textit{temporally similar wards}. For our example, this would mean we would calculate the temporal profile of Italian restaurants for each of the $N$ similar wards.
  \item Create a representative curve. These $N$ temporal curves serve as the basis of our prediction of the profile of our new venue $v_i$. To create a representative curve from those $N$ profiles, we use each of the profiles as inputs to a Gaussian Process (GP) because of its ability to recognize latent periodic trends. The output from the GP becomes our temporal prediction. 
\end{enumerate}

\subsection{Gaussian Processes model} 
Our k-nearest temporal neighbors algorithm finds temporal profiles that are likely to be similar to the venue of interest. We harness Gaussian Processes to build a regression model to capture the periodic trends in those profiles. GPs are typically employed in time-series tasks because of their flexibility at capturing the complex structures without succumbing to overfitting. GP regression is a Bayesian non-parametric which models a distribution over an infinite set of random variables. A GP model is described by its prior mean and covariance functions. For this work, as is standard, we set the prior mean to zero~\cite{rasmussen2006}. For the analysis of this we use a product of two Radial Basis function kernels as the base kernel functions which define the covariance matrix of the distribution. We use two kernels to describe two types of periodicity in our data, over the course of a week as well as over the course of a day.

Given the periodicity over the course of a day and a week, we posit that Gaussian Processes are able to recognize latent periodic trends in the data. The inputs to our Gaussian Process are the temporal profiles of the similar wards. We then have the GP predict a temporal pattern for an interval of $[0, T]$ where $T$ = 168 for the hourly week's profile. We then compare this prediction to stable temporal profile of the venue of interest. 

\subsection{Evaluation}
In this section, in this section we present an evaluation of our algorithm introducing first a set of baselines as comparators and metrics for comparing the experimental results.

\noindent \textbf{Baselines.}
To evaluate our proposed solution, we compare our results with a number of baseline approaches. For each, the past temporal profiles (i.e., earlier than the venue creation time) are used as features for a GP and the outputs of the GP are the prediction of the characteristic temporal profile of the new venue. The baselines and their descriptions are listed in Table~\ref{predValues}.

\noindent\textbf{Metrics.}
To analyze the accuracy of our prediction, we calculate the NRMSE between the predicted temporal profile and the stable profile for each venue. We first look at the value of NRMSE as we vary the number of neighbors $N$. Our results show $N = 10$ to be the best indicator of temporal similarity of neighbors. This value was chosen for the subsequent analysis presented in this paper.

\textbf{Results.} Using $N = 10$, we calculate the normalized root mean squared error for the output of each algorithm compared to the actual stable curve of each new venue. Table~\ref{predValues} presents a summary of these results. Temporally similar wards using the specific category of the venue proves to be the best predictor of the temporal profile of a new venue. 

\begin{table}[h!]  
\caption{Comparative analysis of different similarity criteria.}
\begin{tabular}{ ccc }
  \hline
  \textbf{Criteria} &\textbf{Description of Criteria} & \textbf{NRMSE} \\ 
  \hline 
  \texttt{TempGen}  &Temporally similar wards, same general category & 1.614 \\ 
  \hline 
  \texttt{TempSpec}   & \textbf{Temporally similar wards, same specific category} & \textbf{1.575}\\ 
  \hline 
  \texttt{Random}   &Random wards &  2.692\\ 
  \hline
  \texttt{SameAll} &  Same ward, all categories & 2.1941 \\
  \hline 
  \texttt{SameGen}   &  Same ward, same general category & 1.884 \\
  \hline
  \texttt{SameSpec}  & Same ward, same specific category &1.760 \\
  \hline
  \texttt{AllAll}  & All wards, all categories& 1.937  \\
  \hline
  \texttt{AllGen} & All wards, same general category & 2.190\\
  \hline
  \texttt{AllSpec} &  All wards, same specific category & 2.028 \\
  \hline 
\end{tabular}
\label{predValues}
\end{table}

\section{On-line Prediction of Mobility Trends at New Venues}
\label{sec:realtime}	
As demonstrated in the previous section, predicting the stable temporal curve of a venue is possible and can become more accurate when temporal information from other venues or areas is selectively transferred. Human mobility patterns, despite being characterized by a high degree of regularity, can change over time. Although the characteristic temporal curve provides an indication of the typical temporal trends, the ability to predict demand in real-time can directly help a shop owner to dynamically approximate demand trends, even during the first few months after a venue has opened. In this section, we aim to forecast the popularity of a venue and predict the success of that venue after it has opened. On a day to day basis, we expect the profile of check-ins to vary, however, the variation in demand on a per month basis is less susceptible to noise and a more reliable indicator of the success of a new venue. 

\subsection{On-line prediction task}  
When working at the granularity of individual venues in an on-line manner, the sparsity of the data becomes 
a real concern. In order to address this issue, for each venue, we aggregate check-ins over the course of a month. Predicting a significant increase or decrease in demand for a venue for the following month is useful information for a shop owner to know as it can inform crucial business decisions and who are unlikely to have historic data for the venue. To inform our predictions we utilize locality, the past demand trends of venues in our neighborhood, and temporal similarity, the demand trends of temporally synchronous areas of the city. We next provide a roadmap of how local trends in user mobility can be exploited to forecast alterations of traffic at individual venues in future months.

For each new venue, our prediction task is to estimate the change in demand for each month following the first month of business; then, starting at the first week in which it opened, we aim to predict whether the demand of the venue at its next time step will increase, decrease, or remain stable (i.e., a 3-class prediction task). Remaining stable is defined as remaining within ten percent of the prior value. It is worth noting that the proposed methodology does not depend on the choice of this interval. With each subsequent month since a venue has opened, we have more data with which we can better understand the venue. 

We inform our predictions by selectively using temporal information from other areas of the city. A number of different criteria were used when selecting these inputs, which are listed in Table~\ref{aucvalues}. For example, when using the ``history" of a venue, for each month after the venue opens, we train a GP on the demand for each month. We then predict the relative demand at each progressive month, continuously learning from the previous months (i.e., month 1 to 3 would be used in the training dataset when predicting month 4). This real-time prediction methodology harnesses the Bayesian nature of GPs and their ability to predict and react to anomalies in the data (i.e., if there is a sharp peak at a given time step $t$ the GP uses that insight when predicting $t + 1$). This mirrors situations in which the demand of a venue can sharply increase because of events on a given day.  

\subsection{Evaluation of the on-line prediction task} 
We predict the relative demand curve in real-time for one week for each of our 305 new venues. Table~\ref{aucvalues} presents the AUC value when predicting the relative demand curve after the training phase. 
 
Our baseline is to train the model on the history of the new venue after the first month from it's opening. We examine the use of temporally similar wards as a predictor. We use as our prediction the history of 10 wards because this value provided the optimal representation of similarity. We also examined as inputs venues in wards that have the same general and specific category as the venue of interest. Our results show that locality does have an impact on predictions as the predictions using venues in the same ward are higher than using venues in all wards throughout the city. Further, we see that temporal similarity can also be used to improve predictions; this may be because similar venues could have synchronous peaks in demand following a similar impact from real-world events. 

\begin{table*}[t]
\caption{AUC values of the real-time prediction with a varying number of months of training data.}
\begin{tabular}{ cccccccc } 
 \hline 
			 			& 2 & 3 & 4 & 5 & 6 \\
 \hline
 \texttt{History}                		& 0.6748 & 0.6697 & 0.6853 & 0.7286 & 0.7278 \\
 \hline
 \texttt{TempGen} & 0.7507 & 0.7820 & 0.7691 & 0.7824 & 0.7903 \\
 \hline
 \texttt{TempSpec}  & 0.7729 & 0.7804 & 0.7829 & 0.7991 & 0.8104 \\
 \hline
 \texttt{Random} 		& 0.5102 & 0.5185 & 0.5248 & 0.5682 & 0.5993 \\
 \hline 
 \texttt{SameAll} 	 		& 0.7149 & 0.7310 & 0.7382 & 0.7349 & 0.7352 \\
 \hline 	
 \texttt{SameGen} 		& 0.7403 & 0.7481 & 0.7592 & 0.7480 & 0.7791 \\
 \hline
 \texttt{SameSpec}  	& \bf{0.7859} & \bf{0.7915} & \bf{0.7981} & \bf{0.8216} & \bf{0.8221} \\
 \hline
 \texttt{AllAll} 	 		& 0.6812 & 0.6892 & 0.6489 & 0.6893 & 0.6832 \\
 \hline
 \texttt{AllGen} 	 		& 0.6824 & 0.6853 & 0.6935 & 0.7088 & 0.7129 \\
 \hline 
 \texttt{AllSpec}     		& 0.7201 & 0.7209 & 0.7403 & 0.7459 & 0.7402 \\
 \hline
\end{tabular}
\label{aucvalues}
\end{table*}

\section{Discussion}
The evaluation results discussed in the previous section have shed new light on the temporal
dynamics of user activity in location-based services. 

At neighborhood level, we have seen that areas that are far from each other can be synchronized with regards to their temporal activities.  Moreover, the temporal frequencies of such activities tend to be stationary over certain periods of time due to regularities in human mobility patterns. We exploited this information to predict the temporal popularity profiles of newly established venues in Section~\ref{sec:evaluation}, essentially transferring 
information from the level of an urban region to that of a specific venue. This form of analytics can provide new insights to new business owners who can plan supplies and staffing in their facilities during the cold start 
period of a new opening. Beyond retail venues, the idea can be expanded to other types of places,
such as parks or outdoor spaces. Predicting how urban spaces are used over time can improve planning, including the design of schedules for their maintenance or police them. Despite the regularity patterns observed in human mobility, variations over time will exist 
due to social events or unforeseen circumstances such as travel disruptions. These can result in 
historically unexpected increases or decreases in mobility flows towards a venue. 

To examine whether such variations can be captured on a venue level we experimented with an on-line
prediction task, where the goal was to predict relative changes with respect to historic patterns 
of a venue. This is a challenging task from the point of view of data sparsity. Even very
popular venues in location-based services will have only a handful of check-ins observed in a small time
window. We have demonstrated that it is possible to pick up trends in this setting using a Gaussian Process
model trained on data inputs from recent mobile user activity at nearby venues.

\section{Conclusion and Future Work} 
In this work, we have investigated the prediction of the temporal dynamics of newly established venues using the check-in data of millions of Foursquare users. We have also introduced the concept of \textit{temporally similar areas} in a city, areas that share patterns in the movement of people to different types of venues within those areas. 

We have shown that the characteristic temporal curve of a new venue provides valuable insight for new shop owners who can use that information to better inform supply purchases, opening hours, and demand. Characteristic curves can also support the design of location-based technologies. Additionally, our models help to demonstrate how a particular venue influences the overall temporal profile of the neighborhood it is located in. This knowledge can help design more interpretable models and build urban applications that are aware of the behavioral choices made by citizens on a local level, rather than those that treat population dynamics as a blind optimization task.

We next plan to study venues with different demand characteristics - not only the most popular venues. Less popular venues may attract different demographics or present unconventional temporal properties. Additionally, future work includes the analysis of cities around the world in order to understand their regularity and temporal trends through the application of the methodology discussed in this paper. The framework proposed in this paper can be applied to different cities, since it is not based on any assumptions regarding the spatial and urban context. Indeed, in order to apply the framework, a different model has to be trained, potentially with different values of k and the hyperparameters of the Gaussian Process. However, in order to select these values, it would be sufficient to follow the methodology described in this paper.


\section*{Declarations}
\begin{backmatter}

\section*{Abbreviations}
LBSNs, Online Location-based Social Networks; JSD, Jensen-Shannon divergence; GP, Gaussian Process; NRMSE, normalized root mean squared error.

\section*{Availability of data and material}
The Foursquare dataset used in this work has been shared by the company through an official agreement with the University of Cambridge that we have no authority to redistribute it.

\section*{Competing interests}
The authors declare that they have no competing interests.

\section*{Funding}
This work was supported through the Gates Cambridge Trust and partially supported by the Alan Turing Institute under the EPSRC grant EP/N510129/1. The work was also supported by the EPSRC through Grant GALE (EP/K019392). 

\section*{Authors' contributions}
All authors discussed and designed the experiments as well as contributing to the write up of the paper. KD carried out the computational tasks, analyzed the data and prepared the figures. All authors read and approved the final manuscript.

\section*{Acknowledgements}
We thank Foursquare for supporting this research by providing the dataset employed in the analysis.

\theendnotes

\bibliographystyle{bmc-mathphys} 
\bibliography{Bibliography}   



\end{backmatter}

 \listoffigures

\end{document}